\documentclass[a4paper,11pt]{article}
\usepackage[utf8]{inputenc}
\DeclareUnicodeCharacter{202F}{\,}
\usepackage{color}
\usepackage{cite}
\usepackage{hyperref}
\usepackage{amsmath}
\usepackage{graphicx}

\usepackage[top=0.4in, bottom=0.7in, left=0.60in, right=0.5in]{geometry}

\begin{document}
\title{{\Large $\gamma, J/\psi$  production through ISR process in electron-positron annihilation at B-factories}}
\author{Shashank Bhatnagar}
\maketitle \small{1. Department of Physics, University Institute of Sciences, Chandigarh University, Mohali-140413, India\\}

\begin{abstract}
\normalsize{We study the Initial state radiation (ISR) process, $e^- e^+ \rightarrow^*\gamma\rightarrow \gamma +J/\psi$  observed at high luminosity $e^- e^+$ colliders such as B-factories, with the ISR photon being emitted through the initial electron or positron, prior to their annihilation. The emitted $J/\Psi$ is produced through the photon fragmentation process, with the produced $J/\psi$ in turn decaying into a pair of leptons, $\mu^-+\mu^+$ through the leptonic decay process, $J/\Psi\rightarrow \gamma^*\rightarrow \gamma \mu^- \mu^+$. The cross section for the $J/\Psi$ production, and the branching ratio for the leptonic decay of $J/\Psi\rightarrow \mu^- \mu^+$, both calculated in the framework of Bethe-Salpeter equation are used to work out the cross section for the complete process, $e^- e^+ \rightarrow \gamma +J/\psi\rightarrow \gamma+\mu^- +\mu^+$. Our results are compared with the BaBar data, and other models. This study is then extended to production and decays of $\Upsilon(1S)$ through the ISR process.}
\end{abstract}

\section{Introduction}
Successful experiments at high-luminosity $e^- e^+$ colliders, such as the B-factories have ushered in a new era for studying electron-positron 
annihilation into hadrons at low energies. This progress was achieved using a novel technique known as Initial State Radiation (ISR)\cite{eidelman11,ablikim16,aubert04,cleo06,kedr10}, often referred to as the radiative return method. ISR occurs when a photon is emitted from the incoming electron or positron prior to annihilation, thereby effectively lowering the energy available for producing final states. The application of ISR and the associated radiative return mechanism has become a powerful tool in hadronic physics at low and intermediate energies. This technique enables precise measurements of hadronic cross sections and the $R$-ratio, offering insights into reaction mechanisms, uncovering intermediate resonances and their decay modes, and aiding the identification of new, highly excited mesonic states. Although first discussed theoretically in the 1960s and 1970s, ISR became a practical experimental tool with the development of advanced Monte Carlo generators like EVA-PHOKHARA \cite{eidelman11}, which provided high precision simulations for ISR processes, and have been used in BaBar and Belle experiments at B-factories.

Modern detectors at these colliders, which have accumulated unprecedented integrated luminosities, have enabled ISR-based studies to rival direct  experiments. This approach has yielded a wealth of new data on the cross sections for $e^- e^+$ annihilation into hadrons. Further, valuable information on the particles with mass of about a few GeV has been obtained, primarily on excited vector mesons with radial and/or orbital excitations, their decay channels were studied, with many decays observed for the first time.

The production of heavy quarkonium in electron-positron  annihilation also provides an excellent framework for studying the interplay between perturbative and non-perturbative aspects of Quantum Chromodynamics (QCD). Among the mechanisms facilitating these studies, ISR\cite{eidelman99,chang05,aubert04,aubert07,eidelman11} is of particular significance, as it enables access to intermediate energy regimes even when the center-of-mass (C.M.) energy of the collider is fixed. The ISR-induced production of  final states has attracted significant interest due to its importance in probing QCD dynamics and the properties of heavy quarkonia. As a bound state of $c\bar{c}$, $J/\Psi$ serves as a key probe for understanding strong interactions. Its production, especially through ISR, involves contributions from both perturbative QCD calculations and non-perturbative aspects of quarkonium formation, which are effectively described within frameworks such as Non-Relativistic QCD (NRQCD). In this study, we focus on the production of heavy quarkonium through the process $e^- e^+\rightarrow \gamma+V$ (where V denotes vector quarkonia such as $J/\Psi$ or $\Upsilon(1S)$, or their radially excited states), followed by its leptonic decay to $\mu^- \mu^+$ at B-factories. Now, the ISR process under investigation can occur through virtual photon or Z-boson exchange, where the virtual photon contribution overwhelmingly dominates the Z-boson contribution\cite{chang05}. Thus in this work, we consider diagrams mediated by a virtual photon, and perform the calculations within the framework of the Bethe-Salpeter equation using the leading order diagrams, emphasizing the essential dynamics of ISR-driven quarkonium production.

The ISR processes have been theoretically studied earlier by \cite{eidelman99} at $\Upsilon(4S)$ center of mass energy. At a C.M. energy of 10.6 GeV, the ISR mechanism is particularly effective for studying quarkonium production, as the reduced energy in the quark-antiquark system approaches the $J/\Psi$ mass, resulting in enhanced production rates due to resonance effects and a greater overlap with the quarkonium wavefunction. And since the photon is massless, the kinematics are particularly favorable for studying processes such as $J/\Psi$ production at energies like 10.6 GeV, where the reduced energy corresponds closely to the mass of the $J/\Psi$.

Thus this work presents a comprehensive theoretical analysis of the  production process via ISR in $e^- e^+$ annihilation. By systematically calculating the cross section, we incorporate the effects of ISR and the underlying dynamics of $J/\Psi$ formation. Our predictions are compared with available experimental data, serving as benchmarks for future studies. This analysis contributes to the broader understanding of heavy quarkonium production and underscores the critical role of ISR in $e^- e^+$ collider experiments.

An exclusive quarkonium production process in electron-positron annihilation is dominated by the color-singlet channel, in which a produced $Q\bar{Q}$ pair has the same quantum numbers as that of a final-state quarkonium. Now in  NRQCD, the cross section is factorized into a short distance part (creation of heavy quark pair), and the long distance part (formation of quarkonium), where the short distance component can be calculated perturbatively in strong interaction coupling constant, $\alpha_s$, and is relativistically covariant. However the long distance component is the non-perturbative part, which should incorporate relativistic corrections. In this connection we wish to mention that for the hard exclusive process, $e^- e^+\rightarrow J/\psi + \eta_c$, there was a large discrepancy between leading order NRQCD\cite{braaten} prediction, and Belle\cite{belle04} data, which could only be resolved when both radiative corrections and relativistic corrections were taken simultaneously.

In the present work, we evaluate the exclusive process, $e^- e^+\rightarrow \gamma^*+V$,($V=J/\Psi, \Psi(2S), \Upsilon(1S), \Upsilon(2S)$) within the framework of the Bethe-Salpeter Equation (BSE), a dynamical approach rooted in quantum field theory with a fully relativistic character. The BSE framework not only incorporates the relativistic effects of quark spins but also provides a covariant description of the internal motion of constituent quarks within the hadron. These features enable significant contributions to the cross section at leading order, as demonstrated in our previous studies on processes such as  $e^- e^+\rightarrow \gamma + H$ ($H=\chi_{c0}, \chi_{c1}, \eta_c)$\cite{bhatnagar24}, $e^- e^+\rightarrow \gamma^*\gamma^*\rightarrow J/\Psi+J/\Psi$, and  $e^- e^+\rightarrow \gamma^*\gamma^*\rightarrow \eta_c+\eta_c$\cite{bhatnagar25} at 10.6 GeV. And given the large energy scale of these production processes, the lowest-order perturbative QCD calculations with an $\alpha_s$ expansion, are expected to suffice, as higher-order QCD corrections are likely negligible. Consequently, this study does not consider higher-order diagrams in $\alpha_s$ or corrections from pure electromagnetic interactions, as their contributions are anticipated to be insignificant. Thus in the present work dealing with $\gamma+V$, production at 10.6 GeV, we will study these processes at leading order in BSE under covariant instantaneous ansatz. Such leading order calculations for such ISR processes were also done earlier in \cite{chang05}. Now since in our work we calculate  the cross section for $e^- e^+\rightarrow \gamma+J/\Psi$ at the leading order, this corresponds closely to the Born cross section.  We have thus compared our theoretical results with the Born cross section estimated by BaBar in \cite{aubert04}. 

The paper is organized as follows: In Section 2, we have briefly given the BSE framework for quarkonium bound state under Covariant Instantaneous Ansatz. In Section 3, we have studied the process, 
$e^- e^+\rightarrow \gamma + V$ (where $V=J/\Psi,\Psi(2S), \Upsilon(1S)$ and $\Upsilon(2S)$), followed by the cross section calculation for the full process, $e^- e^+\rightarrow \gamma + V\rightarrow \gamma+\mu^-+\mu^+$. Section 4 is devoted to Discussions.

\section{BSE Framework for Quarkonium bound state}
A 4D Bethe-Salpeter equation (BSE) for quark-antiquark ($Q\bar{Q}$) bound state system can be expressed as,
\begin{equation}
\Psi(P,q)=S_F(p_1)i\int \frac{d^{4}q'}{(2\pi)^{4}}K(q,q')\Psi(P,q')S_F(p_2)
\end{equation}

where the two particles' momenta are $p_1$ and $p_2$, with the hadron's internal momentum being $q$ and the external momentum being $P$ and mass, $M$. In Eq. (1), the interaction kernel is denoted by $K(q,q')$, and the quark and antiquark's inverse propagators are denoted by $S_{F}^{-1}(\pm p_{1,2})=\pm i{\not}p_{1,2}+ m_{1,2}$. Here we wish to mention that the dressed quark propagators are in principle the solutions to the gap equation \cite{bhagwat06,roberts96,roberts11} and characterized by a momentum-dependent mass function, $m(p)$. However, previous studies have demonstrated that employing constituent quark masses for heavy quarks (c, b) is a good approximation \cite{chang10,wang05}. This approach also provides a theoretical basis for the use of constituent quark masses in potential models for heavy quarks. Furthermore, recent BSE calculations \cite{shi20,guo19,chang10,vaishali21,wang22,wang16,wang05,he21} have employed constituent quark masses in heavy-quark propagators. Therefore, in this work, we adopt constituent quark masses for c and b quarks in the propagators. However for further works, we intend to use dressed quark propagators, for which the framework would have to be modified.

The above equation can be reduced to 3D form by applying the Covariant Instantaneous Ansatz (CIA)-a Lorentz-invariant generalization of the Instantaneous Approximation (IA) to the BS kernel, $K(q,q')$, where $K(q,q')=K(\widehat{q},\widehat{q}')$. The BS kernel then depends on the hadron's internal momentum component,  $\widehat{q}_\mu= q_\mu- \frac{q.P}{P^2}P_\mu$ which is a 3-D variable, that is orthogonal to the external hadron momentum, i.e. $\widehat{q}.P=0$, whereas $\sigma P_\mu=\frac{q.P}{P^2}P_\mu$ is the longitudinal component of $q$, and $d^4q=d^3\widehat{q}Md\sigma$ is the 4-D volume element in momentum space. A series of steps detailed in \cite{bhatnagar24} leads to the definition of the 4D BS wave function

\begin{eqnarray}
&&\nonumber \Psi(P, q)=S_1(p_1)\Gamma(\hat q)S_2(-p_2),\\&&
\nonumber \Gamma(\hat{q})=\int \frac{d^3\hat{q}'}{(2\pi)^3}K(\hat{q},\hat{q}')\psi(\hat{q}'),\\&&
\psi(\hat{q}')=i\int\frac{Md\sigma'}{2\pi} \Psi(P,q'),
\end{eqnarray}

with the hadron-quark vertex function, $\Gamma(\hat{q})$ sandwiched between two quark propagators, and is employed for calculation of transition amplitudes for various processes. Also $\psi(\hat{q})$ is the 3D BS wave function obtained by integrating the 4D BS wave function, $\Psi(P,q')$ over the longitudinal component, $Md\sigma'$ of internal hadron momentum as shown in previous equation. Also the 3D reduction of the BSE leads to four Salpeter equations\cite{eshete19},

\begin{eqnarray}
 &&\nonumber(M-\omega_1-\omega_2)\psi^{++}(\hat{q})=\Lambda_{1}^{+}(\hat{q})\Gamma(\hat{q})\Lambda_{2}^{+}(\hat{q})\\&&
   \nonumber(M+\omega_1+\omega_2)\psi^{--}(\hat{q})=-\Lambda_{1}^{-}(\hat{q})\Gamma(\hat{q})\Lambda_{2}^{-}(\hat{q})\\&&
\nonumber \psi^{+-}(\hat{q})=0.\\&&
 \psi^{-+}(\hat{q})=0\label{fw5}
\end{eqnarray}

where $\psi^{\pm\pm}(\hat q)= \Lambda^\pm_{1}(\hat q)\frac{{\not}P}{M}\psi(\hat q)\frac{{\not}P}{M}\Lambda^\pm_{2}(\hat q)$, and the projection operators, $\Lambda^\pm_{j}(\hat q)=\frac{1}{2\omega_j}\bigg[\frac{{\not}P}{M}\omega_j\pm J(j)(im_j+{\not}\hat q)\bigg]$, with $J(j)=(-1)^{j+1},~~j=1,2$. It is to be noted that both the 3D Salpeter equations as well as the 4D hadron-quark vertex function depend on the variable, $\hat{q}^2=q^2-(q.P)^2/P^2$, which is positive definite on hadron mass shell ($P^2=-M^2$) over the entire time-like region of the 4D space. Also $\hat{q}^2$ is a Lorentz-invariant variable (consequently $|\hat{q}|=\sqrt{ q^2-(q.P)^2/P^2}$, which is the length  of the 3D vector $\hat{q}$ is also positive definite, and is a Lorentz-invariant variable\cite{lmenew18,bhatnagar20}), which leads to Lorentz-covariance of the 3D forms of transition amplitudes as well as the 3D Salpeter equations, due to their dependence on $\hat{q}^2$. This increases the applicability of BSE under CIA across different energy scales.

The 3D Salpeter equations lead to mass spectral equations that are used not only for the determination of mass spectrum of ground and excited states of $0^{++}, 0^{-+}, 1^{--}, 1^{++}$ and $1^{+-}$ quarkonium\cite{eshete19}, but also the determination of their radial wave functions, that are in turn employed for transition amplitude calculations of various processes, such as leptonic decays, M1 and E1 radiative decays, two-photon decays (for details, see \cite{lmenew18,eshete19,vaishali21a,vaishali21, bhatnagar20,vaishali24}). Besides the above low energy transitions, we have also studied the hadron production processes \cite{bhatnagar24,bhatnagar25, monika23} in $e^- e^+$ annihilation at $\sqrt{s}$= 10.6GeV on lines of \cite{shi20}, where the results of our calculations of these processes were in good agreement with experiment and with other models. Thus we have tested our BSE framework for processes across different energy scales in an integrated framework, with a common set of input parameters for processes studied from mass spectrum to various transition amplitudes. 

The interaction kernel in BSE is taken to be vector type to make connection with the QCD degrees of freedom. It is one-gluon-exchange like as regards the colour and spin dependence, and thus has a general structure \cite{eshete19,bhatnagar24},

\begin{eqnarray}
&&\nonumber K(\hat{q},\hat{q}')=(\frac{1}{2}\lambda_1.\frac{1}{2}\lambda_2)\gamma_{\mu}\times \gamma_{\mu} V(\hat{q}, \hat{q}')\\&&
\nonumber V(\hat{q},\hat{q}')= \frac{4\pi\alpha_s(M^2)}{(\hat{q}-\hat{q}')^2}
 +\frac{3}{4}\omega^2_{q\bar q}\int d^3r\bigg(\kappa r^2-\frac{C_0}{\omega_0^2}\bigg)e^{i(\hat q-\hat q').\vec r}\equiv V_{OGE}(\hat{q},\hat{q}')+V_{conf.}(\hat{q},\hat{q}'),\\&&
\nonumber \omega_{q\bar{q}}^2=2m\omega_0^2\alpha_s(M^2),\\&&
\alpha_s(M^2)=\frac{12\pi}{(33-2f)}[Log(M^2/\Lambda^2)]^{-1}.
\end{eqnarray}

Here the running coupling constant $\alpha_s(M^2)$ has the values: 0.27522 (for $J/\Psi$), 0.25874 (for $\Psi(2S)$), 0.2125 (for $\Upsilon(1S)$) and 0.20937 (for $\Upsilon(2S)$). The use of one-gluon-exchange kernel in our BSE framework preserves the connection with the gauge invariance. The scalar part of the kernel, $V(\hat{q},\hat{q}')\equiv V(\hat{k})$, with $\hat{k}_{\mu}=\hat{q}_{\mu}-\hat{q}_{\mu}'$, and $\hat{k}_{\mu}=k_{\mu}-\frac{k.P}{P^2}P_{\mu}$ is transverse to $P_{\mu}$, and $\hat{k}^2 \geq 0$ and is a four-scalar over the entire 4-dimensional space. Here $V(\hat{k})$ is written as a sum of the perturbative (one-gluon-exchange) part and the non-perturbative (confinement) part as in previous equation, while $ \kappa=(1+4\hat m_1\hat m_2A_0M^2r^2)^{-\frac{1}{2}}$.  The presence of running coupling constant, $\alpha_s$ in flavour dependent spring constant, $\omega_{q\bar{q}}^2$ provides an explicit QCD motivation to the BSE kernel. To do numerical calculations, we need to fix the scale in strong coupling constant, $\alpha_s(Q^2)$. We have thus fixed this scale as the meson mass, M. Thus we use $\alpha_s(M^2)$ in the interaction kernel on lines of \cite{hluf16,eshete19,wang16}. It is seen that the algebraic form of the confining potential ensures a smooth transition from nearly harmonic (for $c\bar{u}$) to almost linear (for $b\bar{b}$)as is believed to be true for QCD (see\cite{hluf16,bhatnagar24,eshete19} for details regarding the nature of the confining potential).

Further, the interaction kernel in our Bethe-Salpeter Equation (BSE) framework is chosen to be of vector type to establish a direct connection with the fundamental degrees of freedom in QCD. Specifically, adopting a vector form of the kernel as $(\gamma_{\mu}\times \gamma_{\mu})$ allows us to relate the scalar part of the BSE kernel, $V(\hat{k})$ to the scalar structure of the gluon propagator\cite{bhatnagar24}, encompassing both perturbative (one-gluon exchange) and non-perturbative regimes. This connection is expressed through $K(\hat{q},\hat{q}')=\gamma_{\mu}D_{\mu\nu}(\hat{k})\gamma_{\nu}$, where the gluon propagator in Landau gauge\footnote{The Landau gauge is particularly useful in such studies, as it yields a purely transverse propagator, simplifies analytical structures, and is favored in non-perturbative QCD analyses such as Dyson–Schwinger and lattice studies—especially for investigating confinement.} is given by $D_{\mu\nu}(\hat{k})= D(\hat{k})(\delta_{\mu\nu}-\frac{\hat{k}_{\mu}\hat{k}_{\nu}}{\hat{k}^2})$, with $D(\hat{k})=D'(\hat{k})+D''(\hat{k})$. Here $D'(\hat{k})= V_{OGE}$ represents the exact structure of the one-gluon-exchange interaction in the perturbative regime, while the confining part of the interaction,  $V_{Conf.}=D''(\hat{k})$ is modeled as the scalar component of the gluon propagator in the infrared domain. This treatment reflects the fact that the infrared behavior of the gluon propagator is crucial for understanding confinement. Such phenomenological forms of the gluon propagator in the infrared have also been employed by \cite{hilger15}, since QCD has yet to yield a closed-form expression for the BSE kernel suitable for calculating the confining interaction.

Now in our earlier work \cite{hluf16}, the input parameters were chosen to reproduce the mass spectrum of equal-mass heavy quarkonia ($c\bar{c}$ and $b\bar{b}$) within the Bethe-Salpeter Equation framework under the Covariant Instantaneous Ansatz (CIA). In the present manuscript, we refer to this parameter set as Set A: $C_0=0.21,~ \omega_0=0.15 GeV, ~A_0=0.01,~ \Lambda_{QCD}=0.200 GeV,~ m_c=1.490 GeV,~ m_b=4.690 GeV$. We note that the value of $m_b=5.070 GeV$ used in Ref. \cite{hluf16} has since been revised. The corrected value adopted in the present analysis is 4.690 GeV, which not only offers a more accurate representation of bottom quark dynamics, but also leads to improved consistency with bottomonium observables within our Bethe-Salpeter Equation framework.

Subsequently, in order to study mass spectra and radiative transitions involving heavy-light systems such as $c\bar{u},~c\bar{s},~c\bar{b}$ as well as $c\bar{c}$, 
we reparametrized the model and obtained a new Parameter Set B \cite{eshete19}:  $C_0=0.69,~ \omega_0=0.22 GeV,~ A_0=0.01,~ \Lambda_{QCD}=0.250 GeV,~m_c=1.490 GeV,~ m_b=4.690GeV$ (along with input masses of $u,~d,~s$ quarks). This reparametrization was essential to incorporate flavor asymmetry through the inclusion of unequal-mass kinematics, which arises from the asymmetric momentum distribution between the heavy and light quarks. The differing dynamical behavior in heavy-light mesons plays a crucial role in shaping the structure of the Bethe-Salpeter wave function and the resulting meson observables. However the Set B was not intended or tuned to describe bottomonium states. In fact, using Set B for $b\bar{b}$ systems leads to a poor match with experimental decay widths for $\Upsilon(1S)$, while Set A yields results closer to data, though as seen from Table 2, there remains discrepancy in leptonic decay widths of $\Upsilon$ in comparison to $J/\Psi$, which we have addressed in the Discussions. Therefore, for the present study of leptonic decay widths of vector quarkonia, we employ Set A, which is specifically optimized for equal-mass $Q\bar{Q}$.

However for sake of comparison we have listed the decay widths and branching ratios for $V\rightarrow \mu\mu$~ (where $V=J/\Psi,~ \Psi(2S),~\Upsilon(1S),~\Upsilon(2S))$ in Table 3 for both parameter sets A and B. The table clearly shows that while both sets yield reasonable results for $J/\Psi$, the set A reproduces the $\Upsilon(1S)$ reasonably well, thereby validating its use for equal mass quarkonia. And as established in the leptonic decay analysis, Parameter Set A provides more accurate results for both $J/\Psi$ and $\Upsilon(1S)$ mesons. Therefore, we compute and present cross-section results for the ISR processes $e^- e^+ \rightarrow\gamma\rightarrow \gamma +J/\psi \rightarrow \gamma \mu^- \mu^+$, and $e^- e^+ \rightarrow\gamma\rightarrow \gamma +\Upsilon \rightarrow \gamma \mu^- \mu^+$ using only Set A.

\section{Cross section for the ISR mediated process  $e^- e^+\rightarrow \gamma + J/\Psi\rightarrow \gamma+\mu^-+\mu^+$}
In this section, we study the process $e^- e^+ \rightarrow\gamma\rightarrow \gamma +J/\psi$, which is an ISR process that proceeds through a single virtual photon since  the  $J/\psi$ has charge parity, $C=-1$. There are two types of leading-order (LO) Feynman diagrams, shown in Figure 1. In Fig.1a, the ISR photon is emitted from the initial positron, before it interacts with the electron thereby producing a virtual photon that subsequently fragments into the quark-anti-quark pair, that evolves into a vector meson. Whereas in Fig.1b, the ISR photon is emitted from the initial electron before it interacts with the positron. In both the diagrams the invariant mass of the virtual photon is equal to the vector meson mass, $M$. Such photon fragmentation diagrams also contribute to double $J/\Psi$ production processes in electron-positron annihilation \cite{braaten03,bodwin03,luchinsky,peskin} studied earlier.

Let $P,q, \epsilon^{\lambda}$ be the external momentum, internal momentum and the polarization vectors of the outgoing vector meson. We take $k$ as the momenta of the virtual photon involved in production of $J/\Psi$, while ($k', \epsilon^{\lambda'}$) as the momentum, and polarization vector of the ISR photon. This process proceeds through an internal electron line. 

\begin{figure}[ht!]
 \centering
 \includegraphics[width=10cm,height=3cm]{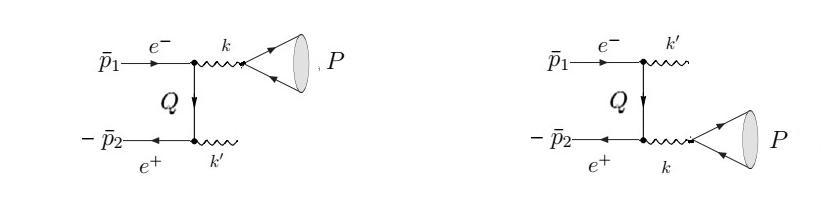}
 \caption{Feynman diagrams involved in $e^- e^+ \rightarrow \gamma+ J/\psi$ at leading order.}
 \end{figure}

The two diagrams in Fig. 1 contribute to this process, whose contributions to amplitude are written down as,

\begin{equation}
{M^1}_{fi}=-i\frac{e^2e_Q}{M^2}[\bar{v}^{(s2)}(\bar{p_2}){\not}\epsilon'[\frac{-i{\not}Q+m_e}{Q^2+m_e^2}]\gamma_{\mu}u^{(s1)}(\bar{p_1})]\int \frac{d^4q}{(2\pi)^4}Tr[\gamma_{\mu}\bar{\Psi}(P,q)],
\end{equation}
and
\begin{equation}
{M^2}_{fi}=-i\frac{e^2e_Q}{M^2}[\bar{v}^{(s2)}(\bar{p_2})\gamma_{\mu}[\frac{-i{\not}Q+m_e}{Q^2+m_e^2}]{\not}\epsilon' u^{(s1)}(\bar{p_1})]\int \frac{d^4q}{(2\pi)^4}Tr[\gamma_{\mu}\bar{\Psi}(P,q)],
\end{equation}

respectively. Here $e_Q=Qe$ where $Q=(+\frac{2}{3};~ -\frac{1}{3})$ for c, and b quarks respectively, while $e=\sqrt{4\pi \alpha_{em}}$. Further $\bar{\Psi}_{V}(P, q)$ is the adjoint 4D BS wave function, and the virtual photon mass is the mass of $J/\Psi$. To reduce the above amplitudes to 3D form, we make use of the Covariant Instantaneous Ansatz (CIA), where the 4D volume element of internal momenta of the produced hadrons can be written as $d^4q=d^3\hat{q} Md\sigma$. We then integrate over the longitudinal component $Md\sigma$ of the internal momenta $q$ of the hadron to obtain,$\int \frac{Md\sigma}{2\pi i}\hat{\Psi}(P,q)=\bar{\Psi}(\hat{q})$. The 3D form of the amplitude is then written as,

\begin{equation}
M^{1}_{fi}=-i\frac{e^2e_Q}{M^2}[\bar{v}^{(s2)}(\bar{p_2}){\not}\epsilon'[\frac{-i{\not}Q+m_e}{Q^2+m_e^2}]\gamma_{\mu}u^{(s1)}(\bar{p_1})]\int \frac{d^3\hat q}{(2\pi)^3}Tr[\gamma_{\mu}\bar{\psi}(\hat q)],
\end{equation}

and
\begin{equation}
M^{2}_{fi}=-i\frac{e^2e_Q}{M^2}[\bar{v}^{(s2)}(\bar{p_2})\gamma_{\mu}[\frac{-i{\not}Q+m_e}{Q^2+m_e^2}]{\not}\epsilon' u^{(s1)}(\bar{p_1})]\int \frac{d^3\hat q}{(2\pi)^3}Tr[\gamma_{\mu}\bar{\psi}(\hat q)],
\end{equation}

\bigskip
\bigskip

If we consider $p_{1,2}$ as the momenta of quarks for the produced hadrons with total momentum $P$, and internal momentum, $q$, the quark momenta are expressed in terms of the internal and total momenta of the produced mesons as:

\begin{equation}
p_1=\frac{1}{2}P + q,~~~~ p_2=\frac{1}{2}P -q.
\end{equation}

Both the incoming and outgoing particle's wave functions are normalised to one particle per unit volume. The expressions above may be factored into the product of leptonic tensor and hadronic tensor: $M^1_{fi}=L_{\mu \nu}^{em} H_{\mu \nu}^{hadronic}$. Now, total amplitude, $M_{fi}=M^{1}_{fi} +M^{2}_{fi}$. However, in order to calculate $M_{fi}$ further, one needs the algebraic forms of 3D adjoint BS wave function, $\bar{\psi}_{V}(\hat{q})$, of vector meson. For this, we first start with the most general form of 4D BS wave function, $\Psi_P(P,q)$ for vector meson $1^{--}$, that is expressed in terms of various Dirac structures in \cite{smith69,alkofer02}. Then, making use of the 3D reduction under Covariant Instantaneous Ansatz, and making use of the fact that $\hat{q}.P=0$, we can write the general decomposition of the instantaneous BS wave function for vector meson of dimensionality $M$ being composed of various Dirac structures that are multiplied with scalar functions $\phi_i(\hat{q})$ as in \cite{wang22}. The amplitudes $\phi_i(\hat{q})$ are all independent. Putting this wave function into the Salpeter equations, Eq.(3) leads to the 3D wave function, $\psi_V(\hat{q})$, whose adjoint wave function, $\bar{\psi}_V(\hat{q})$ is obtained as, $\bar{\psi}_V(\hat{q})=\gamma_4 \psi^\dag_V(\hat{q})\gamma_4$:

\begin{equation}
\bar{\psi}_V(\hat{q})=N_V\bigg[iM{\not}\epsilon+(\hat{q}.\epsilon)\frac{M(m_1+m_2)}{\omega_1\omega_2+m_1m_2-\hat{q}^2}+{\not}{P}{\not}{\epsilon}+i\frac{\omega_1+\omega_2}{2(\omega_1m_2+m_1\omega_2)}({\not}P{\not}\epsilon {\not}\hat{q}+\hat{q}.\epsilon)\bigg]\phi_V(\hat{q})
\end{equation}

which for the two equal mass mesons reduces to,

\begin{equation} \bar{\psi}_V(\hat{q})=N_V\bigg[iM{\not}\epsilon+(\hat{q}.\epsilon)\frac{M}{m}+{\not}{P}{\not}{\epsilon}+\frac{i}{2m}{\not}\hat{q}{\not}\epsilon{\not}P+\frac{i}{2m}(\hat{q}.\epsilon){\not}P\bigg]\phi_V(\hat{q}),
\end{equation}

where $N_V$ is the 4D BS normalizer of the final vector meson, obtained through the current conservation condition,

\begin{equation}\label{46}
2iP_\mu=\int \frac{d^{4}q}{(2\pi)^{4}}
\mbox{Tr}\left\{\overline{\Psi}(P,q)\left[\frac{\partial}{\partial
P_\mu}S_{F}^{-1}(p_1)\right]\Psi(P,q)S_{F}^{-1}(-p_2)\right\} +
(1\rightleftharpoons2).
\end{equation}

Here $\Psi(P,q)$ is the 4D BS wave function, while $\overline{\Psi}(P,q)=\gamma_4 \Psi(P,q)^{\dag}\gamma_4$ is the adjoint BS wave function. We first carry 
out derivatives of inverse quark propagators of constituent quarks with respect to total hadron momentum $P_{\mu}$.  Making use of the fact that $S_{F}^{-1}(p_1)=m+i{\not}p_1$, where $p_1=\frac{1}{2}P+q$, the trace part of the above equation can be written as: $TR= Tr[\overline{\Psi}(P,q)(\frac{i}{2}\gamma_{\mu})\Psi(P,q)S_{F}^{-1}(-p_2)]$, where $\Psi(P,q)=S_F(p_1)\Gamma(\hat{q})S_F(-p_2)$, and $\overline{\Psi}(P,q)=S_F(-p_2)\overline{\Gamma}(\hat{q})S_F(p_1)$. We then make use of the fact that we can express the quark propagators as\cite{bhatnagar20,vaishali21}:

\begin{eqnarray}
&&\nonumber S_F(p_1)=\frac{\Lambda_{1}^{+}(\hat{q})}{M\sigma+\frac{1}{2}M-\omega+i\epsilon}+\frac{\Lambda_{1}^{-}(\hat{q})}{M\sigma+\frac{1}{2}M+\omega-i\epsilon}\\&&
S_F(-p_2)=\frac{-\Lambda_{2}^{+}(\hat{q})}{-M\sigma+\frac{1}{2}M-\omega+i\epsilon}+\frac{-\Lambda_{2}^{-}(\hat{q})}{-M\sigma+\frac{1}{2}M+\omega-i\epsilon}
\end{eqnarray}

Further since the $++$ components of 3D wave functions in Salpeter equations contribute much more than $--$ components to any calculation, while the contributions from $+-$ and $-+$ components is zero from the last two Salpeter equations in Eqs. (3), we take only the $++$ components \cite{hluf17}, and write the Trace product as:

\begin{equation}
[TR]\approx Tr\bigg[\frac{\overline{\Gamma}(\hat{q})\Lambda_{1}^{+}(\hat{q})\gamma_{\mu}\Lambda_{1}^{+}(\hat{q})\Gamma(\hat{q})\Lambda_{2}^{+}(\hat{q})}{[M\sigma+\frac{1}{2}M-\omega+i\epsilon][-M\sigma+\frac{1}{2}M-\omega+i\epsilon]^2}\bigg]
\end{equation}

Now taking the 4D volume element, $d^4 q= d^3\hat{q}Md\sigma$, and following the procedure in \cite{bhatnagar20, vaishali21}, we multiply the above Trace relation from the left by,
 $\frac{{\not}P}{M}\frac{{\not}P}{M}=-1=\frac{{\not}P}{M}(\Lambda_{2}^{+}(\hat{q})+\Lambda_{2}^{-}(\hat{q}))$.  We write Eq.(12) as \footnote{Here it is to be noted that $\Lambda_{2}^{-}(\hat{q})$ multiplied on the left will lead to an additional term $ {\not}P\Lambda_{2}^{-}(\hat{q})\overline{\Gamma}(\hat{q})\Lambda_{1}^{+}(\hat{q})\gamma_{\mu}\Lambda_{1}^{+}(\hat{q})\Gamma(\hat{q})\Lambda_{2}^{+}(\hat{q})$, which vanishes on account of it being expressed in terms of $\overline{\psi}^{-+}$, which is zero from the last two Salpeter equations in Eqs.(3).}: 
\begin{equation}
 2iP_{\mu}=\frac{1}{M^2}\int \frac{d^3 \hat{q}}{(2\pi)^3}\int \frac{Md\sigma}{2\pi i}Tr\bigg[{\not}P\frac{\Lambda_{2}^{+}(\hat{q})\overline{\Gamma}(\hat{q})\Lambda_{1}^{+}(\hat{q})\gamma_{\mu}\Lambda_{1}^{+}(\hat{q})\Gamma(\hat{q})\Lambda_{2}^{+}(\hat{q})}{[\sigma-(\frac{1}{2}-\frac{\omega}{M}+i\epsilon)]^2(\sigma-(-\frac{1}{2}+\frac{\omega}{M}-i\epsilon))}\bigg]+(1\rightleftharpoons 2).
 \end{equation}

Making use of the 3D Salpeter equations, $(M-2\omega)\psi^{++}(\hat{q})=\Lambda_{1}^{+}(\hat{q})\Gamma(\hat{q})\Lambda_{2}^{+}(\hat{q})$, and $(M-2\omega)\overline{\psi}^{++}(\hat{q})=\Lambda_{2}^{+}(\hat{q})\overline{\Gamma}(\hat{q})\Lambda_{1}^{+}(\hat{q})$

the above equation can be written as:
\begin{equation}
2i P_{\mu}=\frac{1}{M^2}\int \frac{d^3\hat{q}}{(2\pi)^3}(M-2\omega)^2 Tr[\alpha {\not}P\overline{\psi}^{++}(\hat{q})\gamma_{\mu}{\psi}^{++}(\hat{q})]+(1\rightleftharpoons 2),
\end{equation}
 where $\psi^{++}(\hat q)= \Lambda^{+}_{1}(\hat q)\frac{{\not}P}{M}\psi(\hat q)\frac{{\not}P}{M}\Lambda^{+}_{2}(\hat q)$, $\overline{\psi}^{++}(\hat q)= \Lambda^{+}_{2}(\hat q)\frac{{\not}P}{M}\psi(\hat q)\frac{{\not}P}{M}\Lambda^{+}_{1}(\hat q)$, and $\alpha$ is the result of $d\sigma$-integration over the poles of the quark propagators, which lie above and below the real axis) in the complex $\sigma$-plane.

It can be checked that the result of integration over $d\sigma$, whether we close the contour below or above the real $\sigma$-axis is the same. And the factor $(M-2\omega)^2$ in the numerator of Eq.(16) exactly cancels with the denominator of the result of contour integration, $\alpha$. We put the 3D BS wave function, $\psi(\hat{q})$ from Eq.(11) into the expressions for $\overline{\psi}^{++}(\hat{q})$, and $\psi^{++}(\hat{q})$ in Eq.(16) . Evaluating trace over products of gamma matrices, following usual steps, we then express the above equation in terms of the integration variables $\hat{q}$. We then multiply both sides of the above equation by $P_{\mu}$. We further make use of the orthogonality conditions, $P.\hat{q}=0$, and $P.\epsilon=0$ to obtain\cite{hluf16}:

\begin{equation}
N_{V}^{-2}=16Mm\int \frac{d^3\hat{q}}{(2\pi)^3}\frac{\hat{q}^2}{\omega^3}\phi_{V}^2(\hat{q}).
\end{equation}

Then numerical integration over the variable $\hat{q}$ is finally performed to extract out the numerical results for BS normalizer, $N_V$ for different equal mass vector mesons. 
However, it is observed that all Dirac structures in BS wave function of a hadron do not contribute equally. And the most dominant Dirac structure, $i\gamma.\epsilon$ contributes maximum to vector meson observables, as seen from earlier studies in \cite{munczek92,sauli12,shi06,jorge11,bhatnagar14}, which is valid not only for meson ground states, but for their excited states as well. And in the calculations studied in this work, the contribution comes only from the most dominant Dirac structure $i\gamma.\epsilon$, while the other Dirac structures do not contribute.

Now the 3D radial wave function $\phi_V(\hat{q})$ satisfies the mass spectral equation, that resembles the equation of a 3D harmonic oscillator (see Refs.\cite{eshete19} for details). Expressing the laplacian operator in this equation in spherical polar co-ordinates, we express mass spectral equation as a  partial differential equation in $\phi_V(\hat{q})$ in terms of orbital angular momentum quantum number, $l = 0,1,2…$ corresponding to $S, P, D,...$ wave states. The solutions of this 3D harmonic oscillator equation is obtained using a power series method, where the eigen values of this equation are: $E_N=2\beta^2 (N +3/2)$, where the principal quantum number, $N=2n+l$; with $n= 0,1,2,…$. We take $l=0 (S$ state) and $l=2 (D$ state)  for vector ($1^{--}$) mesons. Thus, solutions of mass spectral equation\cite{eshete19} not only leads to spectrum of quarkonia, but also their radial wave functions, that are wave functions with definite quantum numbers, $N$ and $l$.

The analytic structures of the unnormalized 3D radial wave functions \cite{bhatnagar20} for $1^{--}$ mesons are\cite{eshete19}:
\begin{eqnarray}\label{wavefunc}
&&\nonumber \phi_V(1S,\hat q)= e^{-\frac{\hat{q}^2}{2{\beta_V}^2}};\\&&
\phi_V(2S,\hat{q})=(1-\frac{2\hat{q}^2}{3{\beta_V}^2})e^{-\frac{\hat{q}^2}{2{\beta_V}^2}},
\end{eqnarray}

where $\beta_V$ is the inverse range parameter with its numerical values for ground and excited states of $J/\Psi$, and $\Upsilon$ listed in Table 1.

\bigskip

Now regarding amplitude for the process, $e^- e^+\rightarrow \gamma+J/\Psi$: after trace evaluation over $\gamma$-matrices in Eqs.(7-8), the amplitude $M_{fi}$ for photon fragmentation diagrams, with contributions from both the Direct and the Exchange diagrams is then written in the centre of mass frame (where $\vec{\bar{p_1}}=- \vec{\bar{p_2}}$, $E_1=E_2 (=E)$, and $\sqrt{s}=2E$) as,

\begin{eqnarray}
&&\nonumber M_{fi}=\frac{e^2 e_Q}{M^2}\int \frac{d^3\hat{q}}{(2\pi)^3}\phi_V(\hat{q})\\&&
\nonumber \bigg[(4M) \{[\bar{v}^{(s2)}(\bar{p_2}){\not}\epsilon'\frac{(-i{\not}Q+m_e)}{Q^2+m_e^2}{\not}\epsilon u^{(s1)}(\bar{p_1})]+[\bar{v}^{(s2)}(\bar{p_2}){\not}\epsilon\frac{(-i{\not}Q+m_e)}{Q^2+m_e^2}{\not}\epsilon' u^{s1}(\bar{p_1})] \}+\\&&
\frac{4}{m}(\hat{q}.\epsilon)\{[\bar{v}^{(s2)}(\bar{p_2}){\not}\epsilon'\frac{(-i{\not}Q+m_e)}{Q^2+m_e^2}{\not}P u^{s1}(\bar{p_1})]+ [\bar{v}^{(s2)}(\bar{p_2}){\not}P\frac{(-i{\not}Q+m_e)}{Q^2+m_e^2}{\not}\epsilon' u^{(s1)}(\bar{p_1})]\bigg].
\end{eqnarray}

We now introduce the 3D integrals, $G$  over the 3D wave function of the outgoing vector meson,

\begin{equation}
 G=\int \frac{d^3\hat{q}}{(2\pi)^3}\phi_V(\hat{q}),
\end{equation}

It is to be noted that the $d^3\hat{q}$ integration over the $(\hat{q}.\epsilon)$ multiplied by the symmetric S-wave function $\phi_V(\hat{q})$ should be zero. This is due to the reason that $\hat{q}$ is a directional vector and its integration inherently eliminates all angular dependencies, and no residual angular terms should persist in the final expressions. Thus only the terms in the first line of Eq.(19) contributes to $M_{fi}$, while the contribution of second line involving terms $(\hat{q}.\epsilon)$ is 0. $M_{fi}$ thus reduces to:

\begin{equation}
M_{fi}=\frac{e^2 e_Q}{M^2}G (4M) \bigg[[\bar{v}^{(s2)}(\bar{p_2}){\not}\epsilon'\frac{(-i{\not}Q+m_e)}{Q^2+m_e^2}{\not}\epsilon u^{(s1)}(\bar{p_1})]+[\bar{v}^{(s2)}(\bar{p_2}){\not}\epsilon\frac{(-i{\not}Q+m_e)}{Q^2+m_e^2}{\not}\epsilon' u^{s1}(\bar{p_1})]\bigg].
\end{equation}

Also it can be checked that the terms on the left and right in Eq.(21) contribute equally to $M_{fi}$. Further as mentioned above, both these terms receive contribution only from the most dominant Dirac structure, $iM{\not}\epsilon$ in the structure of the 3D BS wave function, $\phi_V(\hat{q})$ in Eq.(11).

Further, the importance of the 3D integral, $G$ in Eq.(20) is brought out by the fact that in the process $e^- e^+\rightarrow \gamma + J/\Psi$, followed by $J/\Psi\rightarrow \mu^-+\mu^+$ studied in this work, the integrals involved in $M_{fi}$ expressions in Eqs.(7-8) for $e^- e^+\rightarrow \gamma +J/\Psi$  are exactly identical to the integrals appearing in RHS of Eq.(31) for the leptonic decay, $J/\Psi\rightarrow \mu^-\mu^+$. From a physical perspective: in Fig. 1, $J/\Psi$ is generated through the photon vector current, while in Eqs. (31-32), $J/\Psi$ is annihilated via the photon vector current. Thus both the $J/\Psi$ terms should correspond to their decay constants, that are in turn expressed in terms of the 3D integral, $G$ in Eq.(20), that have a major role to play in the calculation of the cross section. Their numerical values along with the values of inverse range parameter $\beta_V$, BS normalizers $N_V$, for ground and excited states of V mesons are given in Table 1.

\begin{table}[hhhhh]
  \begin{center}
  \begin{tabular}{p{6.3cm} p{2cm} p{2.2cm} p{2.2cm}}
  \hline
 Process                                   &$\beta_V$    &$N_V$  &   $G$ \\
  \hline
  $e^- e^+\rightarrow \gamma J/\psi(1S)$   &  0.4824    & 8.200 & 0.00734\\
   $e^- e^+\rightarrow \gamma \psi(2S)$     & 0.4783    &7.472  &-0.00671\\
    $e^- e^+\rightarrow \gamma \Upsilon(1S)$& 0.6743    &5.7645   & 0.0195 \\
   $e^- e^+\rightarrow \gamma \Upsilon(2S)$ & 0.6698    &4.769  &-0.01901 \\
                                       
     \hline
  \end{tabular}
\caption{Numerical values of the inverse range parameter, $\beta_V$ (in $GeV$), BS normalizer, $N_V$ (in $GeV^{-2}$), and the radial integral $G $ (in $GeV^3$) for four different processes calculated in this work using the parameter Set A \cite{hluf16}:  $C_0=0.21,~ \omega_0=0.15 GeV,~A_0=0.01,~\Lambda_{QCD}=0.200 GeV,~~ m_c=1.490 GeV,~ m_b=4.690 GeV$.}
\end{center}
\end{table}

We now evaluate the spin averaged invariant amplitude modulus squared, $|\bar{M}_{fi}|^2=\frac{1}{4}\sum_{s1,s2,\lambda,\lambda'}{M^\dag}_{fi}M_{fi}$, for which we must now average over the initial spin states of $e^-$ and $e^+$ and sum over the polarisation states of the final vector meson. We make use of the normalization relation, $\Sigma_{\lambda} \epsilon_{\mu}^{\lambda} \epsilon_{\nu}^{\lambda} = \frac{1}{3}(\delta_{\mu\nu} + \frac{P_{\mu}P_{\nu}}{M^2})$ for the polarisation vector of the vector meson. The  electromagnetic coupling constant ($\alpha_{em}=\frac{e^2}{4\pi}$), $m$ is the charm quark mass and $m_e$  is the electron mass. Now taking $Q$ to be the momentum of the internal fermionic line in the Feynman diagram, we can express, $Q=-\bar{p}_2+k'$, where the emitted photon momentum, $k'=(\overrightarrow{k}', i|\overrightarrow{k}'|)$. Since in the center of mass frame, $\overrightarrow{k}'=-\overrightarrow{P}$, and $|\overrightarrow{k}'|=|\overrightarrow{P}|$, we can write the square of the momentum, $Q^2=-m_{e}^2-2|\overrightarrow{p}_1||\overrightarrow{P}|Cos\theta+2E|\overrightarrow{P}|$, where for incident electron, its energy and momentum are taken to be nearly equal ($|\bar{p}_1|\approx E=\frac{\sqrt{s}}{2}$), where $(\bar{p_1}+\bar{p_2})^2=-s$.   The energy and momentum of final meson are, $E'=\frac{s+M^2}{2\sqrt{s}}$, and $|\vec{P}'|=\frac{s-M^2}{2\sqrt{s}}$. Further in this center of mass frame, taking $\theta$ to be the angle between the incident beam direction and $J/\Psi$, the dot products of various momenta can be expressed as: 

\begin{eqnarray}
&&\nonumber \bar{p}_1.Q=\frac{s}{2}-m_{e}^2-|\overrightarrow{p}_1||\overrightarrow{P}|Cos\theta-|\overrightarrow{P}|E,\\&&
\nonumber \bar{p}_2.Q= m_{e}^2+|\overrightarrow{p}_1||\overrightarrow{P}|Cos\theta-E|\overrightarrow{P}|\\&&
\nonumber \bar{p}_1.P=|\overrightarrow{p}_1||\overrightarrow{P}|Cos\theta-EE',\\&&
\nonumber \bar{p}_2.P=-|\overrightarrow{p}_1||\overrightarrow{P}|Cos\theta-EE',\\&&
\nonumber P.Q= |\overrightarrow{p}_1||\overrightarrow{P}|Cos\theta+EE'-|\overrightarrow{P}|^2-E'|\overrightarrow{P}|,\\&&
\end{eqnarray}

The expression for $|\bar{M}_{fi}|^2$ is then written as

\begin{equation}
|\bar{M}_{fi}|^2= N_{V}^2 [\Omega];
\end{equation}

where
\begin{eqnarray}
&&\nonumber [\Omega] = \frac{32\, e^4\, e_Q^2\, G^2}{3 M^4 \left( s - s Cos\theta + M^2 (1 +Cos\theta) \right)^2}\times \\&&
\nonumber \bigg[M^6 \left( 3 - 3 Cos\theta - 5 Cos^2\theta + Cos^3\theta \right) \\&&
\nonumber - s^2 (1 - Cos\theta) \left( s (1 -Cos\theta) Cos\theta + m_e^2 (1 + 5 Cos\theta) \right) \\&&
\nonumber + M^4 \left( m_e^2 (-11 + 24 Cos\theta - 5 Cos^2\theta) + s (6 + Cos\theta + 12 Cos^2\theta - 3 Cos^3\theta) \right) \\&&
+ M^2 \left( -8 m_e^4 + 2 m_e^2 s (-3 - 14 Cos\theta + 5 Cos^2\theta)+ 3 s^2 (1 + Cos\theta - 3 Cos^2\theta + Cos^3\theta) \right)\bigg]    
\end{eqnarray}

In center-of-mass frame, the incident flux  reduces to $F=4|\vec{p}|\sqrt{s}$, where $|\vec{p}|$ is the magnitude of momentum of either of the incident particles, and $\sqrt{s}=2E$ is the centre of mass energy of the incident particles. The cross section following a series of steps can be expressed as:

\begin{equation}
\sigma=\frac{1}{32\pi^2 s^{3/2}}|\vec{P'}|\int d\Omega' |\bar{M}_{fi}|^2,
\end{equation}

with $|\vec{P'}|=\frac{s-M^2}{2\sqrt{s}}$ being the momenta of outgoing vector meson.  Now, it is possible to verify that the BS normalizer behaves as $N_V \sim M^{-2}$. The 3D integral has dimensions: $G\sim [M^3]$. Consequently, $|\bar{M}_{fi}|^2 \sim [M^0]$. As a result, the cross sectional formula has dimensions $\sigma \sim [M^{-2}]$.

However, when we calculate the cross section at a fixed center-of-mass energy (say, $\sqrt{s}$= 10.6 GeV), we are evaluating the process at a single, idealized kinematic point, corresponding to production of an on-shell hadron accompanied by a photon with vanishing momentum. These calculations do not incorporate the continuous spectrum of emitted photon energies due to ISR, and therefore do not reflect the full physical reality of the process in an actual collider environment, and are not observable. A physical cross section should always include effects of ISR in $e^- e^+\rightarrow \gamma+ J/\Psi$. A realistic cross section that incorporates the effects of ISR is achieved through the convolution formalism discussed later in this section. We thus use our ISR cross sections so obtained to compare with experimental data and other models in Tables 3 and 4. However to estimate experimental data \cite{aubert04} for $\gamma^*\rightarrow \gamma+ J/\Psi$, we took their ratio of the Born cross section for $\sigma(e^+ e^-\rightarrow \gamma^*\rightarrow \gamma+J/\psi\rightarrow \gamma \mu^- \mu^+)$ to their branching fraction, $B_{\mu \mu}\sim 0.0588$\cite{aubert04} to obtain $36.122\pm0.833$ pb.

Now for the complete process, $e^- e^+\rightarrow \gamma^*\rightarrow \gamma+J/\Psi\rightarrow \gamma+ \mu^- +\mu^+$, the cross section can be expressed as,

\begin{equation}
\sigma_{e^- e^+\rightarrow \gamma+J/\Psi\rightarrow  \gamma\mu^- \mu^+}=\sigma^{ISR}_{e^- e^+\rightarrow \gamma+J/\Psi}\times B_{\mu\mu},
\end{equation}

where, we multiply the ISR corrected convoluted cross section for $e^- e^+\rightarrow \gamma\rightarrow \gamma+J/\Psi$ with the branching fraction, $B_{\mu\mu}$ for the process, $J/\Psi\rightarrow \mu^- \mu^+$, where $B_{\mu\mu}$ is the ratio of the decay width, $\Gamma_{\mu^- \mu^+}$ to the full width, $\Gamma=92.6\pm1.7 keV$\cite{navas24}. Now, the decay width for $J/\Psi\rightarrow \mu^- \mu^+$ can be expressed as,

\begin{equation}
\Gamma_{\mu \mu}=\frac{4\pi\alpha_{em}^2 Q^2 f_{V}^2}{3M}(1-\frac{m_{\mu}^2}{M^2})^\frac{1}{2} (1+2\frac{m_{\mu}^2}{M^2}),
\end{equation}

where $m_{\mu}$ is the muon mass. The leptonic decay constant,  $f_V$  of the vector meson is evaluated from the coupling of the quark loop to the vector current as,

\begin{equation}
f_V M\epsilon_{\mu}(P)= <0 |\bar{Q}\gamma_{\mu} Q|V(P)>,
\end{equation}

and can be evaluated through the quark-loop integral,

\begin{equation}
f_V M\epsilon_{\mu}=\sqrt{3}\int \frac{d^4 q}{(2\pi)^4} Tr[\Psi_V(P.q)\gamma_{\mu}].
\end{equation}

Making use of the 4D volume element, $d^4 q=d^3\hat{q} Md\sigma$, and the fact that  $\int \frac{Md\sigma}{2\pi i}\Psi_V(P,q)= \Psi_V(\hat{q})$, we obtain the 3D reduced form,

\begin{equation}
f_V M\epsilon_{\mu}=\sqrt{3}\int \frac{d^3 q}{(2\pi)^3} Tr[\Psi_V(\hat{q})\gamma_{\mu}].
\end{equation}

With the 3D wave function, $\Psi_V(\hat{q})$ taken from Eq.(11), evaluating trace over the gamma matrices, multiplying both sides of the resulting equation by the polarization vector, $\epsilon_{\mu}$ of vector meson, and making use of the fact that $P.\epsilon=0$, the decay constant can then be expressed as,

\begin{equation}
f_V=4\sqrt{3}N_V\int \frac{d^3 q}{(2\pi)^3}\phi_V(\hat{q}).
\end{equation}

Making use of the radial parts of the 3D wave functions $\phi_V(1S, \hat{q})$ for $J/\Psi$, and $\phi_V(2S,\hat{q})$ for $\Psi(2S)$ in Eq.(18), that are derived through the analytic solutions of mass spectral equations in \cite{eshete19}, we obtain their leptonic decay constants for $f_V$, which are listed in Table 2 along with their corresponding decay widths $\Gamma_{\mu\mu}$ (in keV) and their branching fractions, $B_{\mu \mu}$. For calculation of the Branching fractions $B_{\mu\mu}$, we have made use of the full decay widths $\Gamma_{Total}^{Expt.}$=92.6$\pm$1.7 keV (for $J/\Psi$ ), 293$\pm$9 keV (for $\Psi(2S)$), 54.02$\pm$1.25 keV (for $\Upsilon(1S)$), and 31.98$\pm$2.63 keV (for $\Upsilon(2S))$ from PDG Tables\cite{navas24}.

In Table 2, for sake of comparison, we give the numerical values of decay constants $f_V$ (in GeV), decay widths $\Gamma_{\mu\mu}$ (in keV), and branching fraction $B_{\mu\mu}$ for four different processes calculated in this work using two different parameter sets: Set A \cite{hluf16}:  $C_0=0.21,~ \omega_0=0.15 GeV,~A_0=0.01,~\Lambda_{QCD}=0.200 GeV,~~ m_c=1.490 GeV,~ m_b=4.690 GeV$, and Set B\cite{eshete19}:  $C_0=0.69,~ \omega_0=0.22 GeV,~ A_0=0.01,~ \Lambda_{QCD}=0.250 GeV,~m_c=1.490 GeV,~ m_b=4.690GeV$, where the Set A is optimized for equal mass $Q\bar{Q}$, while set B is optimized only for $c\bar{c}$ and heavy-light charmonium mesons. 

\begin{table}[hhhhh]
  \begin{center}
  \begin{tabular}{p{2.5cm} p{1cm} p{2cm} p{2cm} p{1.5cm} p{2.5cm} p{3cm}}
  \hline
  Process                           & Set        & $f_V$ [GeV]& $\Gamma_{\mu\mu}$ [keV]&  $B_{\mu\mu}$& $\Gamma_{\mu\mu}^{Expt.}$[keV] & $B_{\mu\mu}^{Expt.}$ \\
  \hline
$J/\psi\rightarrow \mu\mu$          & A     & 0.4169      &5.5680               & 0.0601       & $5.519\pm 0.101$    &$0.0596\pm 0.0003$  \\
                                    &B      &0.4299       &5.9210                &0.0639         &                      &                     \\
$\psi(2S)\rightarrow \mu\mu $       & A     &0.3470      &3.2524                  & 0.0111        & $2.344\pm 0.1758$     &$0.008\pm 0.0006$   \\  
                                    & B     &0.3806       &3.9033                &0.0133         &                      &                     \\
$\Upsilon(1S)\rightarrow \mu\mu$   & A      & 0.7789      & 1.5907                &0.0294         & $1.3397\pm 0.0216$    & $0.0248\pm 0.0004$ \\
                                   & B      & 0.8801      & 2.0307               &0.0376         &                       &                     \\
$\Upsilon(2S)\rightarrow \mu\mu$   & A      & 0.6282     &0.9766                &0.03053        & $0.6216\pm 0.0507$    & $0.0193\pm 0.0017$   \\ 
                                   & B      &0.7298       &1.3178                &0.0412         &                        &                     \\  
  \hline
  \end{tabular}
\caption{Numerical values of Decay constants $f_V$ (in GeV), Decay widths $\Gamma_{\mu\mu}$ (in keV), and Branching fraction $B_{\mu\mu}$ for four different processes calculated in this work using two different parameter sets: Set A \cite{hluf16}, and Set B\cite{eshete19}. The experimental values for $\Gamma_{\mu\mu}^{Exp.}$\cite{navas24}, and $B_{\mu\mu}^{Exp.}$\cite{navas24} are given in last two columns.}
\end{center}
\end{table}

However, it can be seen from Table 2, that the Set A, originally fitted to the spectroscopy of both $c\bar{c}$ and $b\bar{b}$ states, is more consistent parameter set for predicting both charmonium and bottomonium leptonic decay widths and branching ratios, though with smaller discrepancy for decay widths for $\Upsilon$ than the set B (that was not optimised for $b\bar{b}$ states). Thus we have made use of parameter Set A in the cross-section calculations for ISR processes involving $J/\Psi$ and $\Upsilon(1S)$. The values of electronic decay widths calculated in our work are: $\Gamma_{ee}$= 5.5681 keV (for $J/\Psi$) and 1.59428 keV(for $\Upsilon(1S)$).

We now discuss the cross section calculation for $e^- e^+\rightarrow \gamma J/\Psi$ that incorporates the effects of ISR through the convolution formalism \cite{kuraev85,eidelman99,aubert04}  mentioned earlier. This approach effectively captures the probability distribution for the emission of collinear photons by the initial $e^- e^+$ pair, allowing the system to radiatively return to the resonance mass of the $J/\Psi$. Thus for the complete process, $e^- e^+\rightarrow \gamma J/\Psi \rightarrow \mu^- \mu^+ $, the cross section for the full chain is the convoluted cross section for $e^- e^+\rightarrow \gamma J/\psi$ times the branching ratios for $V(nS)\rightarrow \mu^- \mu^+$ (where $V=J/\Psi,\Upsilon)$.

In our calculation we use the Bethe–Salpeter Equation (BSE) framework to compute the subprocess cross section, $\hat{\sigma}(s(1 - x))$ from Eqs.(23-25), and the ISR correction is incorporated externally via convolution with $W(s,x)$ that incorporate leading QED corrections such as collinear photon emission from initial leptons, and logarithmic enhancements $\sim \alpha \ln (s/m_e^2)$, by redistributing the center-of-mass energy and allowing access to the resonance over a range of photon energies. Now, although the leading-order radiator function captures the dominant collinear behavior and leading logarithms, it does not include soft-photon exponentiation, virtual loop corrections, or multiple photon emissions. For higher-precision predictions, resummed approaches such as the Kuraev–Fadin structure function formalism \cite{kuraev85} can be employed to include these effects to all orders in $\alpha_{em}$ within the leading-log approximation.

The ISR corrections via radiator functions have also been employed in earlier theoretical studies \cite{eidelman99, aubert04}, and this approach forms the basis for ISR measurements at B-factories \cite{eidelman11, ablikim16, kedr10}, where the Born cross section includes soft-photon ISR corrections but excludes vacuum polarization and multi-photon effects. We thus calculate the ISR-corrected Born cross section using the convolution formalism involving the radiator function $W(s,x)$ \cite{aubert04,kuraev85}. The Born cross section for $J/\Psi$ production process is given as\cite{aubert04}:

\begin{equation}
d\sigma^{ISR}(e^- + e^+\rightarrow \gamma+J/\Psi)= W(s,x)~ dx~ \hat{\sigma}(s(1-x)).
\end{equation}

We thus use the corrected formula given by the convolution integral:

\begin{equation}
\sigma^{ISR}(e^- + e^+\rightarrow \gamma+J/\Psi)=\int dx W(s,x) \hat{\sigma}(s(1-x)),
\end{equation}

where $W(s,x)$ is the ISR radiator function\cite{aubert04}:

\begin{equation}
W(s,x)=\frac{2\alpha_{em}}{\pi x}(2ln \frac{\sqrt{s}}{m_e}-1)(1-x+\frac{x^2}{2}),
\end{equation}

that describes the probability that an ISR electron or positron emits a photon with energy fraction, $x=2E_\gamma/\sqrt{s}$. It also takes into account the logarithmic enhancement due to collinear radiation. It captures the radiation spectrum of photons in $e^- e^+\rightarrow \gamma+V$. Here, $W(s,x)$ involves the fine structure constant, $\alpha_{em}=\frac{e^2}{4\pi}=\frac{1}{137}$ , and depends on the center of mass energy $\sqrt{s}$ of initial electron and positron, photon energy (via $x$) and the electron mass, $m_e$. Further $\hat{\sigma}(s(1-x))$ in Eq.(33) is the cross section evaluated at reduced energy center-of-mass energy, $s'=s(1-x)$ due to ISR photon carrying away the energy. Eqs.(33-34) give a  correct estimate for the radiative photon emission cross section with about $10-20 \% $ precision\cite{eidelman99}.

Thus once we include ISR using the radiator function $W(s,x)$, we are integrating over a range of reduced effective energies $\sqrt{s'}=\sqrt{s(1-x)}$, and the cross section is enhanced near the resonance (i.e. when $\sqrt{s'}\sim M_V$). And since the ISR function peaks at low $x$, and the resonance is sharply peaked at $s'\sim M_{V}^2$, the integral gains sufficient weight near this point leading to enhancement of cross section. This can be visualized from Fig. 2, that gives the 3D plot of the of the full integrand $\sigma(s(1-x))\times W(s,x)$ in Eq.(33), which is a 2-dimensional function of the variables $x~\epsilon~ [0, 1-M^2/s]$ and $Cos\theta~\epsilon~[-1,+1]$ for the processes, $e^- e^+\rightarrow \gamma+J/\Psi$ and $e^- e^+\rightarrow \gamma+\Upsilon$ at $\sqrt{s}$= 10.6 GeV.

\begin{figure}[ht!]
 \centering
 \includegraphics[width=14cm,height=4cm]{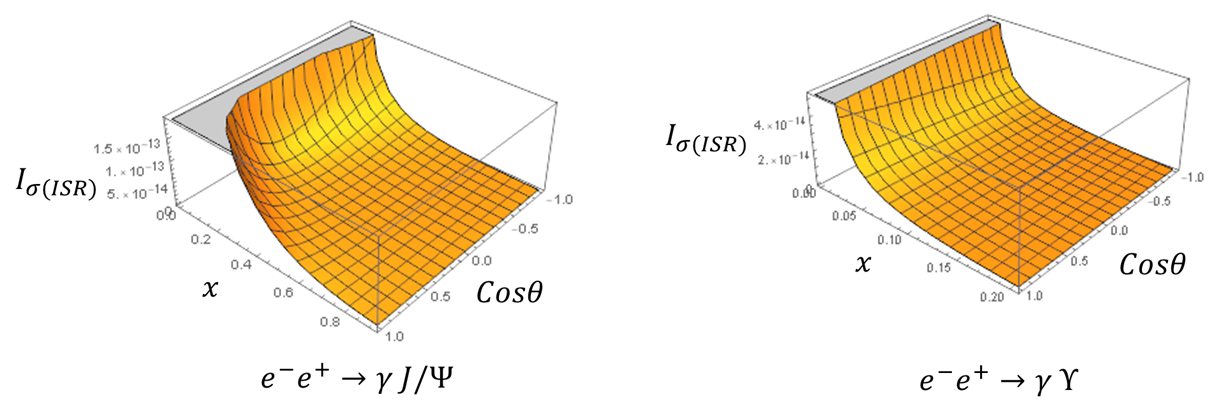}
 \caption{3D plots of the integrand, $I_{\sigma^(ISR)}$ (in barns) of Eq.(33) for the processes $e^- e^+\rightarrow \gamma+J/\Psi$ and $e^- e^+\rightarrow \gamma+\Upsilon$ at $\sqrt{s}$= 10.6 GeV, which is a 2-dimensional function of $x$ and $Cos\theta$.}
 \end{figure}

Here the QED radiator function $W(s,x)$ enhances the integrand in the soft photon region (small $x$) due to its characteristic $1/x$ behaviour that reflects 
the dominance of low-energy ISR photon emissions. Our 3D plots in Fig. 2, as a function of $x$ and $Cos\theta$ shows a pronounced peak in the small-$x$ region, validating this enhancement. We obtain the ISR corrected cross sections, $\sigma^{ISR}(e^- e^+\rightarrow \gamma+J/\Psi)$= 32.94 pb, and $\sigma(e^- e^+\rightarrow \gamma+\Upsilon)$=9.3312 pb as in Table 3.

However, percentage enhancements due to ISR cannot be meaningfully defined for exclusive processes like $e^- e^+ \rightarrow \gamma + J/\psi$,  as there is no experimentally measurable fixed-point cross section for comparison, and thus are typically not reported in literature. The Born-level cross section evaluated at a fixed $\sqrt{s}$ corresponds to a kinematic configuration where the emitted photon carries vanishing momentum, which is not realised. However analogous studies in inclusive processes\cite{shao14} such as: $e^- e^+\rightarrow J/\Psi +X$ at B-factories report $15- 25\%$ enhancements.

To study the behavior of the ISR cross section, $\sigma_{ISR}(s)$, for the process $e^- e^+ \rightarrow \gamma + J/\Psi$, we plotted its variation with $\sqrt{s}$ in the energy range 4 – 10.6 GeV 
in Fig.3. 

\begin{figure}[ht!]
 \centering
 \includegraphics[width=8cm,height=5cm]{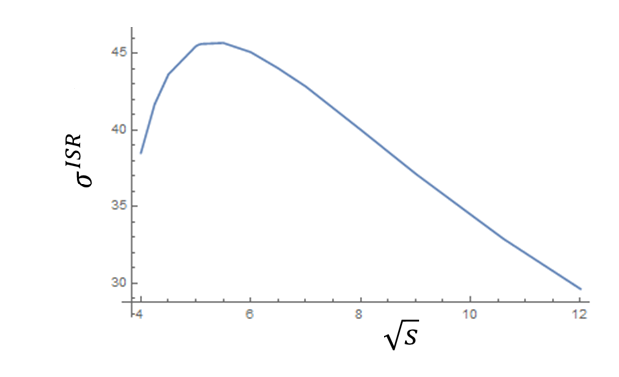}
 \caption{Plot of variation of $\sigma^{ISR}$ (in pb) versus the center of mass energy $\sqrt{s}$  for the process $e^- e^+ \rightarrow \gamma+ J/\Psi$ in the region $\sqrt{s}$= 4- 12 GeV}
 \end{figure}

The resulting curve shows a clear peak around $\sqrt{s} \approx 5.5$–6 GeV, followed by a smooth fall-off at higher energies. This peak arises due to the combined effect of the radiator function $W(s,x)$, which enhances the contribution from events where the effective center-of-mass energy $s(1 - x)$ is near the resonance region, and the large value of the subprocess cross section $\hat{\sigma}(s(1 - x))$ in that domain. As $\sqrt{s}$ increases beyond the resonance region, the ISR-corrected cross section exhibits a power-law suppression. This reflects the expected scaling behavior of the underlying QED subprocess, $\sigma \sim 1/s^n$ with $n > 1$. However, the drop in the cross section as $\sqrt{s}$ decreases from 5 GeV to 4 GeV may be due to phase space suppression at lower energies. Thus even after accounting for ISR, there is insufficient energy to produce an on-shell $J/\Psi$ with a recoiling photon. Thus, the overall trend of $\sigma^{ISR}(s)$ is fully consistent with both QED expectations and physical kinematic constraints. A similar behaviour of ISR cross section with $\sqrt{s}$ is observed in $e^- e^+ \rightarrow \gamma+\Upsilon$.

In Tables 3 and 4, where we give the cross sections $\sigma^{ISR}(e^- e^+\rightarrow \gamma V)$  (through use of radiator function $W(s,x)$), and the resulting $\sigma^{ISR}(e^- e^+\rightarrow \gamma^*\rightarrow \gamma V \rightarrow \gamma \mu^- \mu^+)$ for the full chain, calculated in present work after combining $\sigma^{ISR}(e^- e^+\rightarrow \gamma J/\Psi)$ with corresponding branching ratio $B_{\mu\mu}$ at $\sqrt{s}=10.6 GeV$, where our results are compared with experiment\cite{aubert04}, and other models\cite{eidelman99,chang05}.

\begin{table}[hhhhh]
  \begin{center}
  \begin{tabular}{p{6.5cm} p{2.2cm} p{4cm}p{3cm} p{3cm}}
  \hline
  Process                                  & BSE-CIA&Expt.\cite{aubert04}&\cite{eidelman99}&\cite{chang05} \\
  \hline
  $\sigma^{ISR}(e^- e^+\rightarrow \gamma J/\Psi)$       &32.942   &36.122$\pm$0.833&34.00          & 70    \\
  $\sigma^{ISR}(e^- e^+\rightarrow \gamma \Psi(2S))$     &19.2027  &                 &13.007         &     \\
  $\sigma^{ISR}(e^- e^+\rightarrow \gamma \Upsilon(1S))$  &9.3312    &                &21             &      \\
  $\sigma^{ISR}(e^- e^+\rightarrow \gamma \Upsilon(2S))$ &0.828   &                   &16             &  \\
     \hline
  \end{tabular}
\caption{ISR Corrected Cross sections (in pb) for process, $e^- e^+\rightarrow \gamma^*\rightarrow \gamma V$ (where $V= J/\Psi, \Psi(2S), \Upsilon(1S), \Upsilon(2S))$ calculated in present work with use of radiator function $W(s,x)$ using the parameter set A\cite{hluf16}.}
\end{center}
\end{table}

\begin{table}[hhhhh]
  \begin{center}
  \begin{tabular}{p{7.5cm} p{3.2cm} p{4.3cm}}
  \hline
  Process                                  & BSE-CIA  & Expt.\cite{aubert04}\\
  \hline
  $\sigma^{ISR}(e^- e^+\rightarrow \gamma J/\psi\rightarrow \gamma \mu^- \mu^+)$  &1.9798&2.124$\pm$0.049$\pm$0.047\\
  $\sigma^{ISR}(e^- e^+\rightarrow \gamma \psi(2S)\rightarrow \gamma \mu^- \mu^+)$& 0.2132&        \\
  $\sigma^{ISR}(e^- e^+\rightarrow \gamma \Upsilon (1S)\rightarrow \gamma \mu^- \mu^+)$&0.2753 &        \\
  $\sigma^{ISR}(e^- e^+\rightarrow \gamma \Upsilon(2S)\rightarrow \gamma \mu^- \mu^+)$& 0.0258  &        \\
\hline
\end{tabular}
\caption{ISR corrected Cross sections (in pb) for the enire chain $e^- e^+\rightarrow \gamma^*\rightarrow \gamma V \rightarrow \gamma \mu^- \mu^+$ (where $V= J/\Psi, \Psi(2S), \Upsilon(1S), \Upsilon(2S))$ calculated in present work after combining $\sigma^{ISR}(e^- e^+\rightarrow \gamma J/\Psi)$ with corresponding branching ratio $B_{\mu\mu}$ at $\sqrt{s}=10.6 GeV$ along with experimental data\cite{aubert04} where available.}
\end{center}
\end{table}

\section{Discussion}
We have studied the ISR mediated process, $e^- e^+\rightarrow \gamma^*\rightarrow \gamma+V\rightarrow \mu^- + \mu^+ +\gamma$ ($V$ being the vector quarkonium), where the photon in final state is emitted from the initial electron or positron at center of mass energy, $\sqrt{s}$=10.6 GeV in the framework of Bethe-Salpeter equation under Covariant Instantaneous Ansatz. In this calculation, we have incorporated all the Dirac structures in BS wave of the produced vector meson in contrast to considering only the leading order Dirac structures in some of the works. Our ISR-corrected cross sections for $e^- e^+\rightarrow \gamma^*\rightarrow \gamma+V$, (with $V=J/\Psi(nS); \Upsilon(nS), (n=1,2))$ are given in Table 3 along with results of other models\cite{eidelman99,chang05}. 
We obtain the production cross section $\sigma^{ISR}(e^- e^+\rightarrow \gamma^*\rightarrow \gamma+J/\Psi)$=32.94 pb, which when combined with the branching ratio $B_{\mu\mu}=0.0601$  for $J/\Psi\rightarrow \mu^- \mu^+$, yields the total cross section $\sigma^{ISR}(e^- e^+\rightarrow \gamma^*\rightarrow \gamma+J/\Psi\rightarrow \gamma \mu^- \mu^+)$=1.9798 pb, which is in good agreement with the experimental Born-level cross section reported by BaBar $\sigma^{Born}(e^- e^+\rightarrow \gamma^*\rightarrow \gamma+J/\Psi\rightarrow \gamma \mu^- \mu^+)=2.124\pm0.049\pm0.047$ pb~\cite{aubert04}, where the Born cross section includes soft-photon ISR corrections but excludes vacuum polarization and multi-photon effects.

To evaluate ISR corrected cross sections, we employed a convolution formalism in which the cross section is expressed as an integral over the radiator function $W(s,x)$ derived from QED, that is folded with the reduced-energy Born-level cross section $\sigma (s(1-x))$. The inclusion of $W(s,x)$  enhances the integrand in the soft photon region (small $x$) due to its characteristic $1/x$ behaviour, reflecting the dominance of low-energy ISR photon emissions. Our 3D plot of the full integrand $\sigma(s(1-x))\times W(s,x)$ as a function of $x$ and $Cos\theta$ shows a pronounced peak in the small-$x$ region, validating this enhancement. This enhancement arises purely from incorporating soft and collinear photon emission effects through the convolution formalism and does not involve any higher-order loop or vertex corrections. The result is in good agreement with experimental data and underscores the importance of ISR corrections in precision cross section predictions.

As regards leptonic decays, from Table 2, we see significant discrepancy in decay width of $\Upsilon$ in comparison to $J/\Psi$. We wish to mention that the parameters in Set A\cite{hluf16} were obtained by fitting the mass spectral equations using only the confining potential, with the Coulomb (OGE) term omitted \cite{hluf16}. Although QCD corrections are generally more pronounced in the charmonium sector, the relativistic effects (arising from the full Dirac structure of the 4D Bethe–Salpeter wave functions and the 4D BS normalizers), along with the broader spatial spread of the $c\bar{c}$ confining potential, enable a better approximation of its full dynamics. This, in turn, partially compensates for the absence of short- distance (Coulombic) interactions and leads to a reasonable agreement with experimental data for the leptonic decay width of the $J/\Psi$. Further, the parameter fitting naturally absorbs the effect of the missing short-range contributions more effectively in the charmonium system than in the more tightly bound bottomonium states, where short-distance dynamics play a more dominant role.

Thus in case of $b\bar{b}$ systems, the short-range dynamics is more critical, and the stronger impact of the neglected Coulomb interaction (important at shorter distances) due to the more tightly bound $b\bar{b}$ system cannot be as effectively replicated by the confining potential alone. In our present calculation of decay widths, though the numerical values of the inverse range parameter $\beta$ (that is dependent on the input kernel, and contains the entire dynamical information), were fixed using the full interaction kernel ($V_{Coul.}+V_{Conf.}$) to account for the wave function behavior at the origin, however, since the original parameter fit (Set A) lacked the OGE term, the decay width for $\Upsilon$ remains less accurate.

Thus the apparent success in the charmonium sector reflects a tuning of parameters within a limited interaction kernel used to calibrate set A of parameters\cite{hluf16}, rather than a contradiction of QCD expectations. In \cite{hluf16} which led to Set A of parameters, our intention was to demonstrate that the Bethe–Salpeter approach with only confinement can yield reasonable first approximations. We acknowledge this limitation and are currently extending our work by incorporating the one-gluon-exchange (OGE) term in both the mass spectrum and decay widths to systematically improve accuracy in the bottomonium sector- having already studied $c\bar{c}$ and heavy-light charmonium mesons for various $J^{PC}$ states in \cite{eshete19,vaishali21a} (but not $b\bar{b}$) with the incorporation of the full BSE kernel $(V_{Conf.} + V_{OGE})$ to get the parameter set B. This approach\cite{eshete19,vaishali21a} had led to extremely good agreement with experimental data for both mass spectra and leptonic decay constants, including the relativistically sensitive charmonium system. However in these works,\cite{eshete19,vaishali21a}, since we did not study $b\bar{b}$ systems, we are using set A\cite{hluf16} in present work, where we studied both $c\bar{c}$ and $b\bar{b}$, but with only the confinement interaction used to fix the parameters, in spite of fixing the inverse range parameter $\beta$ by taking account of the full interaction (Confining + Coulomb).

It is to be mentioned that the ISR preferentially produces lighter vector mesons because the probability of radiating a high-energy photon (needed to produce a heavier meson) is lower. This is evident from the fact that since $\Upsilon(1S, 2S)$ are much heavier than $(J/\Psi, \Psi)$, their production through ISR is much smaller. This fact is also observed in the difference in cross-sections between $e^- e^+\rightarrow \gamma^*\rightarrow \gamma+J/\Psi\rightarrow \mu^- \mu^+$, and  $e^- e^+\rightarrow \gamma^*\rightarrow \gamma+\Upsilon(1S)\rightarrow \mu^- \mu^+$, which can be attributed to several key factors. First, the production rate of a vector meson via initial-state radiation (ISR) is proportional to the square of the quark charge, giving an enhancement factor of 4 for $J/\Psi (c\bar{c})$ over $\Upsilon(b\bar{b})$. Second, the available phase space for ISR emission favours the production of the lighter 
$J/\Psi$, leading to a higher cross-section. Additionally, the electronic width $\Gamma_{ee}$ of $J/\Psi$ (5.5681 keV) is significantly larger than that of $\Upsilon (1S)$ (1.59428 keV), 
further enhancing its production probability. Finally, the branching fraction for $J/\Psi\rightarrow \mu^- \mu^+ (6.01\%$) is also higher than that of $\Upsilon(1S)\rightarrow \mu^- \mu^+ (2.95\%$). These combined effects explain the higher observed cross-section for the ISR-mediated $e^- e^+\rightarrow \gamma*\rightarrow \gamma+J/\Psi\rightarrow \mu^- \mu^+$ than $e^- e^+\rightarrow \gamma^*\rightarrow \gamma+\Upsilon(1S)\rightarrow \mu^- \mu^+$.

In conclusion, our study provides a comprehensive analysis of ISR-mediated production of vector quarkonia within the framework of the Bethe-Salpeter equation under Covariant Instantaneous Ansatz, incorporating the full Dirac structure of the vector meson wave function. Our results that bring out the observed hierarchy in the ISR production cross-sections through quark charge, phase space considerations, muon decay widths, and branching fractions are consistent with experimental data and other theoretical models. These insights not only affirm the reliability of the BSE framework in handling exclusive processes involving heavy quarkonia, but also highlight its potential for exploring similar other processes for future studies.

\end{document}